\documentclass[oneside,12pt]{article}
\usepackage[pdftex]{graphicx}
\usepackage{indentfirst}
\usepackage{amsmath}
\usepackage{amssymb}
\usepackage{multirow}
\usepackage{appendix}
\usepackage{color}
\usepackage{tabularx}
\usepackage{changepage}
\usepackage{longtable}
\usepackage{geometry}
\usepackage{lscape}
\usepackage[backref]{hyperref}
\usepackage{comment}

\DeclareMathOperator{\arctanh}{arctanh}
\DeclareMathOperator{\arcsinh}{arcsinh}
\DeclareMathOperator{\arccoth}{arccoth}
\DeclareMathOperator{\sech}{sech}

\newcommand{\specialcell}[2][c]{%
  \begin{tabular}[#1]{@{}c@{}}#2\end{tabular}}

\numberwithin{equation}{section}
\title{{ Atlas of Coordinate Charts \\ on the de Sitter Spacetime}}
\author{Gabriel Pascu\\ \vspace{-10pt}
\small Faculty of Physics, West University of Timi\c{s}oara\\  \vspace{-10pt}
\small Vasile P\^{a}rvan Avenue 4, Timi\c{s}oara, 300223, Rom\^{a}nia, EU\\
\small gpascu@physics.uvt.ro
}

\numberwithin{equation}{subsection}

\newcommand{\newsection}[1]{%
\newpage
    \section{#1}
    \addtocontents{lot}{%
       \vspace{10pt} \textbf{\Large Section \thesection: #1 \vspace{10pt}}
    }
}

\begin{document}
\maketitle

\begin{abstract}
The de Sitter manifold admits a wide variety of interesting coordinatizations. The 'atlas' is a compilation of the coordinate charts referenced throughout the literature, and is presented in the form of tables, the starting point being the embedding in a higher-dimensional Minkowski spacetime. The metric tensor and the references where the coordinate frame is discussed or used in applications are noted. Additional information is given for the entries with significant use: a convenient tetrad and the form taken by the Killing vectors in the respective coordinate frame. \\\\
\end{abstract}




\newpage
\listoftables

\newpage

\newsection{Introduction}

This work is an attempt to bring together all the known coordinate frames that have been utilised or mentioned in the literature for de Sitter spacetime. There have been a number of rather thorough reviews in this sense, the most important of which is the one made by Eriksen and Gr{\o}n \cite{eriksen1995sitter}, but also in part by Schmidt \cite{schmidt1993sitter}, Bi{\v{c}}{\'a}k and Krtou{\v{s}} \cite{bicak2005fields} or Spradlin, Strominger and Volovich \cite {spradlin2001houches}.

Here, emphasis is put on the embedding in a higher-dimensional Minkowskian spacetime and the form of the line element as it results from this embedding.
The de Sitter manifold $d\mathbb{S}^{1,n}$ (Lorenzian manifold if dimension $n+1$) can be envisioned as a hyperboloid embedded in an $n+2$- dimensional flat Minkowski maifold $\mathbb{M}^{1,n+1}$, acording to the constraint
$$\eta_{AB}Z^A Z^B=-\frac{1}{\omega^2}$$
where $\eta_{AB}$ is the metric on $\mathbb{M}^{1,n+1}$, $A,B$ are indices that run from $0$ to $n+1$, $\{Z^A\}$ is the standard cartesian coordinate chart on $\mathbb{M}^{1,n+1}$ and $\omega$ is the Hubble constant. In this work, we consider only the case with three spatial dimensions ($n=3$), where the results have more familiar forms, but they can be generalised to arbitrary $n$.

The $d\mathbb{S}$ is parametrized by coordinate charts generically denoted by $\{ x^\mu \}$, where $\mu$ runs from $0$ to $n=3$. Then, the metric on $d\mathbb{S}$ that is inherited through the embedding $Z^A(x^\mu)$ is

$$ g_{\mu \nu}= \eta_{AB} \frac{\partial Z^A}{\partial x^\mu} \frac{\partial Z^B}{\partial x^\nu}$$
At a coordinate change, the metric transforms just like a tensor:

$$g^\prime_{\alpha \beta}= g_{\mu\nu}\frac{\partial x^\mu}{\partial x^\alpha} \frac{\partial x^\nu}{\partial x^\beta}$$

The de Sitter manifold, just as the Minkowski manifold is maximally symmetric (it has the maximum amount of Killing vectors- 10 for the total 4 dimensions). The Killing vectors components on de Sitter can be expressed in terms of the embedding, being inherited also from the embedding space:

$$ k_{AB}^\mu= g^{\mu\nu} \eta_{AC} \eta_{BD} \left( Z^A \frac{\partial Z^B}{\partial x^\nu}- Z^B\frac{\partial Z^A}{\partial x^\nu}\right)$$
Therefore, this is a simple way for computing their forms in various charts, not being necessary to get them by solving the Killing equations:

$$ \nabla_\mu k^a_\nu +\nabla_\nu k^a_\mu=0$$
Also, for a coordinate chart transformation, their components transform as componets of a vector:

$$ k_{AB}^\alpha= k_{AB}^\mu \frac{\partial x^\mu}{\partial x^\alpha}$$
which is a convenient relation for computing the Killing vector components between closely related charts, or those that belong to the same class of coordinates.

 We do not stress too much the explicit transformations between coordinates, as in \cite{eriksen1995sitter}. The transformations between the main classes of coordinates are given there, and they can be combined with the ones at the beginning of each section of the present work, in order to obtain the coordinate transformation from any one chart to another.
 
The work is formatted as tables in order to provide the most clear overview possible.
It is intended to be used as a quick reference and guide to the various coordinatisations of the de Sitter manifold, each used in order to express different properties, or to emphasise different connections to other spacetimes. 

As a convention, all coordinates that have a physical meaning of spatial, or temporal coordinate are expressed in terms of length (dimension of $\frac{1}{\omega}$). Sometimes, in the literature, they are scaled in order to be adimensional: $r_{sc}=\omega r$. We do not differentiate between scaled adimensional coordinates and those which have dimension of length ( we consider the chart to be the "same" one). In some references, the constant $\omega$ is dropped altogether from the line element.

In other cases, coordinates are shifted like $r_{sh}=r+a$ usually with a factor of $a=-\frac{\pi}{2 \omega}$, the domain being shifted from $ r \in \left(0, \frac{\pi}{\omega} \right)$ to $ r_{sh} \in \left(-\frac{\pi}{2\omega}, \frac{\pi}{2\omega} \right)$. This is applied to both temporal coordinates, and radial coordinates. It occurs when for example in the line element $\sin$ can be equivalently replaced with $\cos$. Then $\sin(\omega r)=\cos(\omega r_{sh})$ and $\cos(\omega r)=-\sin(\omega r_{sh})$. "Shifted" coordinates defined in this way are denoted by a prime superscript, and are presented only if they have been identified at least once in the cited literature.

There is also when the coordinates are "rotated" in the sense of the coodinate used being pure imaginary: $r_{rot}=i r$. This has been done even by de Sitter himself, as Schmidt \cite{schmidt1993sitter} carefully points out. Then, $\cos(\omega r)=\cosh (\omega r_{rot})$,  $\cosh(\omega r)=\cos (\omega r_{rot})$ and $\sin (\omega r)=-i\sinh(\omega r_{rot})$, $\sinh (\omega r)=-i\sin(\omega r_{rot})$. We do not consider here coordinates with purely imaginary domain, since they are always reducible to ones with real domain. However, there is one instance of a coordinate with complex domain referenced in Table \ref{ccc}.

While these considerations might seem trivial, the lack of a standard convention, or even nomenclature and notation for the coordinates often leads to confusion.

Most of the charts do not cover the manifold in its entirety. In fact, {\it sensu stricto} there is no global chart on de Siiter space, since the spherically symmetric charts, of the type $\{t,r,\theta,\phi\}$ have two trivial coordinate symmetries at the points corresponding to $\theta=0,\phi=2\pi$, just as there is no global chart on the sphere manifold $\mathbb{S}^2$, the chart $\{\theta, \phi\}$ covering everything, except the two poles. We will use the term "global chart" in the relaxed sense (as most of the physics literature tacitly assumes), meaning a chart with possibly trivial coordinate singularities. However, it can happen that such a chart is global, but not smoothly global (a singular surface is present).

Charts that introduce additonal parameters, such as  Bi{\v{c}}{\'a}k and Krtou{\v{s}}'s "accelerated coordinates" \cite{bicak2005fields} are also not considered.

The charts are grouped in an synthetic way, rather than being classified mathematically. For charts with spherical symmetry (except the ones with at least a null coordinate), the version in spherical coordinates is presented head-to-head with the one in cartesian (or more correctly called "pseudo-cartesian") coordinates, even if that one does not appear in the literature. The transformation is the usual one:

$$\begin{aligned}
x &= r \sin \theta \cos \phi\\
y &= r \sin \theta \sin \phi \\
z &= r \cos \theta
\end{aligned}$$

The spherical line element on the sphere manifold $\mathbb{S}^2$ is denoted $d\Omega_2^2=d\theta^2+\sin^2 \theta d\phi^2$, while the one on $\mathbb{H}^{1,1}$ is $dH_2^2=d\Theta^2+\sinh^2 \Theta d\phi^2$.

The notation used for the coordinates in the embedding space is $Z^A$, where the index $A=0..5$. For the coordinates on the de Sitter hyperboloid, the notation is $x^\mu$, where $\mu=0..4$. The temporal index is denoted by $0$ and the spatial ones by $i,j=1..3$. Note that while $x^0=x_0$, $x^i=-x_i$. Therefore $x^\mu x_\mu=t^2-x^{i2}$ and $x^\mu \partial_\mu=t\partial_0 -x^i \partial_i$.

For the various charts, every distinct coordinate is given a different symbol, sometimes with a subscript or an overline. Unfortunately, since the notation across sources is most always inconsistent, we tried to both keep notation used in some works, but also refrain from using the same notations for different coordinates. 
 
Every quantity is expressed using these symbols, except in the case of the Killing vectors, where they are referenced generically as $\{t,x^i\}$ and their derivatives as $\partial_0$ and $\partial_i$ respectively.


We start with 4 families of rather special charts- that are diagonal when expressed in spherical coordinates. 3 of them are the FLRW families, and one is the static family. Of course, any one coordinate can be transformed arbitrarily, giving rise to an infinite number of coordinate charts pertaining to each family. However, they are almost exclusively found in the literature in forms that have the line element of the metric expressed as a FLRW line element in one of its 6 forms- 3 proper, or 3 conformal \cite{ibison2007conformal} (or similar to that- for the static family). These common charts, presented in Sections 2 and 3 can be neatly arranged in the following table, according to the form taken by the line element expressed in spherical coordinates.

\renewcommand{\tabcolsep}{1.5pt}
 \scriptsize{
\begin{longtable}[!h]{|c|c|c|c|c|c|c|c|c||c|c|c|c|c|c|}  
   \caption{{\bf The 4 families with spherical diagonal line elements}} \\ \hline
         \multicolumn{3}{|c|}{ \multirow{3}{*}{$\begin{aligned} & \\ b&=b(t,r) \\ a&=a(t,r) \\ s_i&=s_i(t,r) \\ \Omega&=\Omega(t,r) \end{aligned}$}} & 	\multicolumn{6}{c||}{proper forms}	& 	\multicolumn{6}{c|}{conformal forms}\\	\cline{4-15}
				
				\multicolumn{3}{|c|}{}		& \multicolumn{2}{c|}{standard}& \multicolumn{2}{c|}{alternate} & \multicolumn{4}{c|}{isotropic} & \multicolumn{2}{c|}{alternate} & \multicolumn{2}{c|}{standard} \\ 
			\multicolumn{3}{|c|}{}	& \multicolumn{2}{c|}{$\begin{aligned}   b^2 \cdot dt^2-a^2 \cdot \\  \cdot (dr^2+ s_1^2 \cdot d\Omega^2_2) \end{aligned}$}& \multicolumn{2}{c|}{$\begin{aligned} & b^2 \cdot dt^2-a^2 \\ & \cdot (s_2^2 \cdot dr^2+ \\ &+ r^2 d\Omega^2_2) \end{aligned}$} & \multicolumn{2}{c||}{$\begin{aligned} & b^2 \cdot dt^2-a^2 \\ & \cdot s_3^2 \cdot (dr^2+ \\ &+ r^2 d\Omega^2_2) \end{aligned}$} & \multicolumn{2}{c|}{$\begin{aligned} \Omega^2 \cdot [dt^2- \\ -s_3^2 \cdot (dr^2 + \\ +r^2 d\Omega_2^2)] \end{aligned}$}  & \multicolumn{2}{c|}{$\begin{aligned} \Omega^2 \cdot [dt^2- \\ -s_2^2 \cdot dr^2 - \\ -r^2 d\Omega_2^2) \end{aligned}$} & \multicolumn{2}{c|}{$\begin{aligned} \Omega^2 \cdot (dt^2-\\- dr^2 -s_1^2 \cdot d\Omega_2^2) \end{aligned}$} \\  \hline

\multirow{8}{*}{\specialcell{Isotropic\\$s_i=s_i(r)$}} & \multirow{6}{*}{\specialcell{FLRW \\ $b=1$ \\$a=a(t)$ \\ $\Omega=\Omega(t)$}} & \multirow{2}{*}{\specialcell{k=+1}} &   \multirow{2}{*}{\hyperref[FLRWk1ps]{$\{t_{+1},r_{+1}\}$}} &  \multirow{2}{*}{$\{t_{+1},r_{+1}^\prime \}$} &  \multicolumn{2}{c|}{\multirow{2}{*}{\hyperref[FLRWk1pa]{$\{t_{+1},\bar{r}_{+1}\}$}}} &  \multicolumn{2}{c||}{\multirow{2}{*}{\hyperref[FLRWk1pi]{$\{t_{+1},\rho_{+1}\}$}}}& \multicolumn{2}{c|}{$\{\eta_{+1},\rho_{+1}\}$} & \multicolumn{2}{c|}{$\{\eta_{+1},\bar{r}_{+1}\}$}& \hyperref[FLRWk1cs]{$\{\eta_{+1},r_{+1}\}$}& $\{\eta_{+1},r_{+1}^\prime\}$ \\ \cline{10-15}
&  & &  & &\multicolumn{2}{c|}{}&\multicolumn{2}{c||}{}&\multicolumn{2}{c|}{$\{\eta_{+1}^\prime,\rho_{+1}\}$}& \multicolumn{2}{c|}{$\{\eta_{+1}^\prime,\bar{r}_{+1}\}$}& \hyperref[FLRWk1cs_a]{$\{\eta_{+1}^\prime,r_{+1}\}$} & \hyperref[FLRWk1cs_b]{$\{\eta_{+1}^\prime,r_{+1}^\prime\}$} \\ \cline{3-15}

& & \multirow{2}{*}{\specialcell{k=0 }}  & \multicolumn{6}{c||}{\multirow{2}{*}{\hyperref[FLRWk0p]{$\{t,r\}$} }} & \multicolumn{6}{c|}{\multirow{2}{*}{\hyperref[FLRWk0c]{$\{\eta,r\}$} }}\\  &  &   &  \multicolumn{6}{c||}{} & \multicolumn{6}{c|}{} \\ \cline{3-15}

&  & \multirow{2}{*}{\specialcell{k=-1}}  &   \multicolumn{2}{c|}{\multirow{2}{*}{\hyperref[FLRWkmps]{$\{t_{-1},r_{-1}\}$}}} & \multicolumn{2}{c|}{\multirow{2}{*}{\hyperref[FLRWkmpa]{$\{t_{-1},\bar{r}_{-1}\}$}}} & \multicolumn{2}{c||}{\multirow{2}{*}{\hyperref[FLRWkmpi]{$\{t_{-1},\rho_{-1}\}$}}} & \multicolumn{2}{c|}{\multirow{2}{*}{$\{t_{-1},\rho_{-1}\}$}} & \multicolumn{2}{c|}{\multirow{2}{*}{$\{t_{-1},\bar{r}_{-1}\}$}} & \multicolumn{2}{c|}{\multirow{2}{*}{\hyperref[FLRWkmcs]{$\{t_{-1},r_{-1}\}$}}} \\  
&  &   &  \multicolumn{2}{c|}{}&\multicolumn{2}{c|}{}&\multicolumn{2}{c||}{}&\multicolumn{2}{c|}{}&\multicolumn{2}{c|}{}&\multicolumn{2}{c|}{}\\ \cline{2-15}

&  \multicolumn{2}{c|}{\multirow{2}{*}{\specialcell{static $b=b(r)$ \\$a=1$, $\Omega=\Omega(r)$}}} &  \multirow{2}{*}{\hyperref[staticf]{$\{t_s,\bar{r}_F\}$}} & \multirow{2}{*}{$\{t_s,\bar{r}_F^\prime\}$}  & \multicolumn{2}{c|}{\multirow{2}{*}{\hyperref[static]{$\{t_s,r_s\}$}}} &   \multicolumn{4}{c|}{\multirow{2}{*}{\hyperref[statici]{$\{t_s,\rho_s\}$} }}   & \multicolumn{2}{c|}{\multirow{2}{*}{\hyperref[statica]{$\{t_s,\bar{r}_s\}$}}} &\multicolumn{2}{c|}{\multirow{2}{*}{\hyperref[staticr]{$\{t_s,r^*\}$}}} \\
&  \multicolumn{2}{c|}{}   &  &  &\multicolumn{2}{c|}{}&\multicolumn{4}{c|}{}&\multicolumn{2}{c|}{} & \multicolumn{2}{c|}{} \\ \hline
\end{longtable} }
\renewcommand{\tabcolsep}{6pt}

\normalsize{}

Further, we mean by 'natural charts'- the charts for which there is an embedding
$$ Z^\mu=\frac{x^\mu}{f(x_\nu x^\nu)}$$ 
All the coordinates that make up the special four families of charts with diagonal line elements in spherical coordinates and the natural charts are defined in the context of the special properties that these charts have. But in fact, coordinates from any chart can be combined with any other, provided no two coordinates have an exclusive dependence on eachother. This is what we mean by the made-up notion of "hybrid" coordinates. Because the metric in these situations is usually more complicated (it can containing even $dtdx$ terms), charts of this type are not usually taken into consideration in applications.

After that, some anisotropic charts are presented, and after that charts with null coordinates, of Eddington-Finkelstein, or Kruskal type- named in reference to their analogues defined on the Schwarschild geometry.

Note: This work is not intended for publication. It emerged as a result of a PhD report of the author, under the supervison of prof. dr. Ion I. Cot\u{a}escu, to whom the author is grateful for guidance, encouragement and useful suggestions.

\newgeometry{hmargin={10mm,10mm}}


\newsection{FLRW charts}
\begin{longtable}[h]{c|c}
   \caption{{\bf FLRW coordinates on de Sitter spacetime}} \\
	         \multicolumn{2}{c}{k=+1 FLRW radii \cite[rel.(5.2),(5.7)]{eriksen1995sitter}} \\
           $r_{+1}$ & $r_{+1}=\frac{1}{\omega} \arcsin (\omega \bar{r}_{+1})=\frac{2}{\omega} \arctan \frac{\omega \rho_{+1}}{2}$ \\
					& $dr_{+1}=\frac{d\bar{r}_{+1}}{\sqrt{1-\omega^2 \bar{r}^2_{+1}}}=\frac{d\rho_{+1}}{1+\frac{\omega^2 \rho_{+1}^2}{4}}$ \\		
					 $\bar{r}_{+1}$  &  $\bar{r}_{+1}=\frac{1}{\omega} \sin (\omega r_{+1})= \frac{\rho_{+1}}{1+\frac{\omega^2 \rho_{+1}^2}{4}}$ \\
					& $d\bar{r}_{+1}= \cos (\omega r_{+1}) dr_{+1}=  4 \frac{1-\frac{\omega^2 \rho_{+1}^2}{4}}{\left( 1+\frac{\omega^2 \rho_{+1}^2}{4} \right)^2}d\rho_{+1} $ \\
					$\rho_{+1}$  & $\rho_{+1}=\frac{2}{\omega} \tan \frac{\omega r_{+1}}{2}= \frac{2}{\omega^2 \bar{r}_{+1}}(1-\sqrt{1-\omega^2 \bar{r}^2_{+1}})$  \\
					& $d\rho_{+1}= \frac{dr_{+1}}{\cos^2 \frac{\omega r_{+1}}{2}}= \frac{2}{\omega^2 \bar{r}_{+1}^2} \left( 1- \frac{1}{\sqrt{1-\omega^2 \bar{r}_{+1}^2}}\right) d\bar{r}_{+1}$ \\
\end{longtable}
	
Note that there are many equivalent, but not so obvious ways to express $t_{-1}$ and $\eta_{-1}$.

\newpage
\begin{longtable}[h]{ccc}
   \caption{{\bf Proper comoving/ spatially flat FLRW chart}}\\ \label{FLRWk0p}
     & cartesian $\{t,\vec{x}\}$ & spherical $\{t,r,\theta,\phi\}$ \\ \hline
      
 def. & $\begin{aligned}
Z^0 &=\frac{1}{\omega}\sinh (\omega t)+ \frac{\omega}{2}e^{\omega t}|\vec{x}|^2 \\
Z^i &=e^{\omega t} x^i \\
Z^4 &=\frac{1}{\omega}\cosh (\omega t)- \frac{\omega}{2}e^{\omega t}|\vec{x}|^2
\end{aligned}$ &
 $\begin{aligned}
Z^0 &=\frac{1}{\omega}\sinh (\omega t)+ \frac{\omega}{2}e^{\omega t}r^2 \\
Z^1 &=e^{\omega t} r \sin \theta \cos \phi \\
Z^2 &=e^{\omega t} r \sin \theta \sin \phi \\
Z^3 &=e^{\omega t} r \cos \theta \\
Z^4 &=\frac{1}{\omega}\cosh (\omega t)- \frac{\omega}{2}e^{\omega t}r^2
\end{aligned}$ \\ 
Ranges of & $t \in (-\infty,+\infty)$ & $t \in (-\infty,+\infty)$ \\
coordinates & $\vec{x} \in \mathbb{R}^3_x$ & $r \in (0,+\infty)$ \\
covers & \multicolumn{2}{c}{region $Z^0+Z^4>0$ (half of the manifold)} \\
Refs. & \cite[p.130]{birrell1984quantum} & \cite[rel. (2.1)]{gautreau1983geodesic} \\
  &  \cite[rel. (13-14)]{spradlin2001houches} & \cite[(A9)]{mottola1985particle}\\
\end{longtable}


\begin{longtable}[h]{ccc}
     \caption{{\bf Conformal spatially flat FLRW chart}}\\ \label{FLRWk0c}
     & cartesian $\{\eta,\vec{x}\}$ & spherical $\{\eta,r,\theta,\phi\}$ \\ \hline
      
 def. & $\begin{aligned}
Z^0 &=-\frac{1-\omega^2(\eta^2-\vec{x}^2)}{2\omega^2 \eta}\\
Z^i &=-\frac{x^i}{\omega \eta} \\
Z^4 &=-\frac{1+\omega^2(\eta^2-\vec{x}^2)}{2\omega^2 \eta}\\
\end{aligned}$ &
 $\begin{aligned}
Z^0 &=-\frac{1-\omega^2(\eta^2-r^2)}{2\omega^2 \eta}\\
Z^1 &=-\frac{r \sin \theta \cos \phi}{\omega \eta} \\
Z^2 &=-\frac{r \sin \theta \sin \phi}{\omega \eta} \\
Z^3 &=-\frac{r \cos \theta}{\omega \eta} \\
Z^4 &=-\frac{1+\omega^2(\eta^2-r^2)}{2\omega^2 \eta}
\end{aligned}$ \\ 

Ranges of & $\eta \in (-\infty,0)$ & $\eta \in (-\infty,0)$ \\
coordinates & $\vec{x} \in \mathbb{R}^3_x$ & $r \in (0,+\infty)$ \\ 
cover & \multicolumn{2}{c}{half of the manifold (same as spatially flat FLRW)} \\ 
Extended & $\eta \in (-\infty,+\infty)$ & $\eta \in (-\infty,+\infty)$ \\
coordinates & $\vec{x} \in \mathbb{R}^3_x$ & $r \in (0,+\infty)$ \\
\end{longtable}

\newpage
\begin{longtable}[ht]{ccc}
 \caption{{\bf  k=+1 proper FLRW chart, standard coordinates}}\\ \label{FLRWk1ps}
     & cartesian $\{t_{+1},\vec{x}_{+1}\}$ & spherical $\{t_{+1},r_{+1},\theta,\phi\}$ \\ \hline
      
 def. & $\begin{aligned}
Z^0 &=\frac{1}{\omega}\sinh(\omega t_{+1}) \\
Z^i &= \frac{1}{\omega  |\vec{x}_{+1}|}\cosh(\omega t_{+1}) \sin (\omega |\vec{x}_{+1}|) x_{+1}^i\\
Z^4 &= \frac{1}{\omega}\cosh(\omega t_{+1}) \cos (\omega |\vec{x}_{+1}|)
\end{aligned}$ &
 $\begin{aligned}
Z^0 &=\frac{1}{\omega}\sinh(\omega t_{+1}) \\
Z^1 &=\frac{1}{\omega}\cosh(\omega t_{+1}) \sin (\omega r_{+1}) \sin \theta \cos \phi \\
Z^2 &=\frac{1}{\omega}\cosh(\omega t_{+1}) \sin (\omega r_{+1}) \sin \theta \sin \phi \\
Z^3 &=\frac{1}{\omega}\cosh(\omega t_{+1}) \sin (\omega r_{+1}) \cos \theta \\
Z^4 &=\frac{1}{\omega}\cosh(\omega t_{+1}) \cos (\omega r_{+1}) 
\end{aligned}$ \\ 

Ranges of & $t_{+1} \in (-\infty,+\infty)$ & $t_{+1} \in (-\infty,+\infty)$ \\
coordinates & $|\vec{x}_{+1}| \in \left[0,\frac{\pi}{\omega}\right)$ & $r_{+1} \in \left[0,\frac{\pi}{\omega}\right)$ \\ 
cover & \multicolumn{2}{c}{all the manifold} \\ 

& \multicolumn{2}{c}{comoving} \\

Refs. &  & \cite[p.42]{griffiths2009exact} \cite[p.124]{hawking1973large} \\
\end{longtable}


\begin{longtable}[ht]{ccc}
 \caption{{\bf  k=+1 proper FLRW chart, alternate coordinates}}\\ \label{FLRWk1pa}
     & cartesian $\{t_{+1},\vec{\bar{x}}_{+1}\}$ & spherical $\{t_{+1},\bar{r}_{+1},\theta,\phi\}$ \\ \hline
      
 def. & $\begin{aligned}
Z^0 &=\frac{1}{\omega}\sinh(\omega t_{+1}) \\
Z^i &=\cosh(\omega t_{+1}) \bar{x}^i_{+1} \\
Z^4 &=\frac{1}{\omega}\cosh(\omega t_{+1}) \sqrt{1-\omega^2 \vec{\bar{x}}^2_{+1}}
\end{aligned}$ &
 $\begin{aligned}
Z^0 &=\frac{1}{\omega}\sinh(\omega t_{+1}) \\
Z^1 &=\cosh(\omega t_{+1}) \bar{r}_{+1} \sin \theta \cos \phi \\
Z^2 &=\cosh(\omega t_{+1}) \bar{r}_{+1} \sin \theta \sin \phi \\
Z^3 &=\cosh(\omega t_{+1}) \bar{r}_{+1} \cos \theta \\
Z^4 &=\frac{1}{\omega}\cosh(\omega t_{+1}) \sqrt{1-\omega^2 \bar{r}^2_{+1}}
\end{aligned}$ \\ 

Ranges of & $t_{+1} \in (-\infty,+\infty)$ & $t_{+1} \in (-\infty,+\infty)$ \\
coordinates & $|\vec{\bar{x}}_{+1}| \in \left[0,\frac{1}{\omega}\right)$ & $\bar{r}_{+1} \in \left[0,\frac{1}{\omega}\right)$ \\ 
cover & \multicolumn{2}{c}{half of the manifold -corresponding to $r_{+1} \in \left[0, \frac{\pi}{2\omega} \right)$} \\

Refs. &  & \cite[rel. (4.1)]{eriksen1995sitter} \\
  
line elem. &  & $ds^2=dt_{+1}^2- \cosh^2(\omega t_{+1}) \left( \frac{d\bar{r}_{+1}^2}{1-\omega^2 \bar{r}_{+1}^2}-\bar{r}_{+1}^2 d\Omega^2_2 \right)$ \\
\end{longtable}


\begin{longtable}[ht]{ccc}
 \caption{{\bf k=+1 proper FLRW chart, isotropic coordinates}}\\ \label{FLRWk1pi}
     & cartesian $\{t_{+1},\vec{\bf{x}}_{+1}\}$ & spherical $\{t_{+1},\rho_{+1},\theta,\phi\}$ \\ \hline
      
 def. & $\begin{aligned}
Z^0 &=\frac{1}{\omega}\sinh(\omega t_{+1}) \\
Z^i &=\cosh(\omega t_{+1}) \frac{\bf{x}^i_{+1}}{1+\frac{\omega^2 \bf{x}_{+1}^2}{4}} \\
Z^4 &=\frac{1}{\omega}\cosh(\omega t_{+1}) \sqrt{\frac{1-\frac{\omega^2 \bf{x}_{+1}^2}{4}}{1+\frac{\omega^2 \bf{x}_{+1}^2}{4}}}
\end{aligned}$ &
 $\begin{aligned}
Z^0 &=\frac{1}{\omega}\sinh(\omega t_{+1}) \\
Z^1 &=\cosh(\omega t_{+1}) \frac{\rho_{+1}}{1+\frac{\omega^2 \rho_{+1}^2}{4}} \sin \theta \cos \phi \\
Z^2 &=\cosh(\omega t_{+1}) \frac{\rho_{+1}}{1+\frac{\omega^2 \rho_{+1}^2}{4}}  \sin \theta \sin \phi \\
Z^3 &=\cosh(\omega t_{+1}) \frac{\rho_{+1}}{1+\frac{\omega^2 \rho_{+1}^2}{4}}  \cos \theta \\
Z^4 &=\frac{1}{\omega}\cosh(\omega t_{+1}) \sqrt{1-\frac{\omega^2 \rho_{+1}^2}{1+\frac{\omega^2 \rho_{+1}^2}{4}}}
\end{aligned}$ \\ 

Ranges of & $t_{+1} \in (-\infty,+\infty)$ & $t_{+1} \in (-\infty,+\infty)$ \\
coordinates & $\vec{\bf{x}}_{+1} \in \mathbb{R}^3$& $\rho_{+1} \in [0,\infty)$ \\ 
cover & \multicolumn{2}{c}{all the manifold} \\ 
& \multicolumn{2}{c}{speed of light is isotropic} \\ 

Refs. &  & \cite[rel. (5.9)]{eriksen1995sitter}\\
\end{longtable}


\begin{longtable}[ht]{ccc}
 \caption{{\bf  k=+1 conformal FLRW chart, standard coordinates}}\\ \label{FLRWk1cs}
     & cartesian $\{\eta_{+1},\vec{x}_{+1}\}$ & spherical $\{\eta_{+1},r_{+1},\theta,\phi\}$ \\ \hline
      
 def. & $\begin{aligned}
Z^0 &=\frac{1}{\omega \tan (\omega \eta_{+1})} \\
Z^i &= \frac{1}{\omega  |\vec{x}_{+1}| \sin (\omega \eta_{+1})} \sin (\omega |\vec{x}_{+1}|) x_{+1}^i\\
Z^4 &= \frac{1}{\omega \sin (\omega \eta_{+1})} \cos (\omega |\vec{x}_{+1}|)
\end{aligned}$ &
 $\begin{aligned}
Z^0 &=\frac{1}{\omega \tan (\omega \eta_{+1})} \\
Z^1 &=\frac{1}{\omega \sin (\omega \eta_{+1})} \sin (\omega r_{+1}) \sin \theta \cos \phi \\
Z^2 &=\frac{1}{\omega \sin (\omega \eta_{+1})} \sin (\omega r_{+1}) \sin \theta \sin \phi \\
Z^3 &=\frac{1}{\omega \sin (\omega \eta_{+1})} \sin (\omega r_{+1}) \cos \theta \\
Z^4 &=\frac{1}{\omega \sin (\omega \eta_{+1})} \cos (\omega r_{+1}) 
\end{aligned}$ \\ 

Ranges of & $\eta_{+1} \in \left[0,\frac{\pi}{\omega}\right)$ & $\eta_{+1} \in \left[0,\frac{\pi}{\omega}\right)$ \\
coordinates & $|\vec{x}_{+1}| \in \left[0,\frac{\pi}{\omega}\right)$ & $r_{+1} \in \left[0,\frac{\pi}{\omega}\right)$ \\ 
cover & \multicolumn{2}{c}{all the manifold} \\ 

& \multicolumn{2}{c}{conformal comoving} \\

Refs. &  & \cite[p.43]{griffiths2009exact} \cite[p.134]{birrell1984quantum} \\
\end{longtable}


\begin{longtable}[ht]{ccc}
 \caption{{\bf  k=+1 conformal FLRW chart, standard alternate coordinates}}\\ \label{FLRWk1cs_a}
     & cartesian $\{\eta_{+1}^\prime,\vec{x}_{+1}\}$ & spherical $\{\eta_{+1}^\prime,r_{+1},\theta,\phi\}$ \\ \hline
      
 def. & $\begin{aligned}
Z^0 &=\frac{1}{\omega} \tan (\omega \eta_{+1}^\prime) \\
Z^i &= \frac{1}{\omega  |\vec{x}_{+1}| \cos (\omega \eta_{+1}^\prime)} \sin (\omega |\vec{x}_{+1}|) x_{+1}^i\\
Z^4 &= \frac{1}{\omega \cos (\omega \eta_{+1}^\prime)} \cos (\omega |\vec{x}_{+1}|)
\end{aligned}$ &
 $\begin{aligned}
Z^0 &=\frac{1}{\omega} \tan (\omega \eta_{+1}^\prime) \\
Z^1 &=\frac{1}{\omega \cos (\omega \eta_{+1}^\prime)} \sin (\omega r_{+1}) \sin \theta \cos \phi \\
Z^2 &=\frac{1}{\omega \cos (\omega \eta_{+1}^\prime)} \sin (\omega r_{+1}) \sin \theta \sin \phi \\
Z^3 &=\frac{1}{\omega \cos (\omega \eta_{+1}^\prime)} \sin (\omega r_{+1}) \cos \theta \\
Z^4 &=\frac{1}{\omega \cos (\omega \eta_{+1}^\prime)} \cos (\omega r_{+1}) 
\end{aligned}$ \\ 

Ranges of & $\eta_{+1}^\prime \in \left[-\frac{\pi}{2\omega},\frac{\pi}{2\omega}\right)$ & $\eta_{+1}^\prime \in \left[-\frac{\pi}{2\omega},\frac{\pi}{2\omega}\right)$ \\
coordinates & $|\vec{x}_{+1}| \in \left[0,\frac{\pi}{\omega}\right)$ & $r_{+1} \in \left[0,\frac{\pi}{\omega}\right)$ \\ 
cover & \multicolumn{2}{c}{all the manifold} \\ 

& \multicolumn{2}{c}{conformal comoving} \\

Refs. &  & \cite[rel. (5.16)]{eriksen1995sitter}  \\
\end{longtable}


\begin{longtable}[ht]{ccc}
 \caption{{\bf  k=+1 conformal FLRW chart, alternate standard coordinates}}\\ \label{FLRWk1cs_b}
     & cartesian $\{\eta_{+1}^\prime,\vec{x}_{+1}^\prime \}$ & spherical $\{\eta_{+1}^\prime,r_{+1}^\prime,\theta,\phi\}$ \\ \hline
      
 def. & $\begin{aligned}
Z^0 &=\frac{1}{\omega} \tan (\omega \eta_{+1}^\prime) \\
Z^i &= \frac{1}{\omega  |\vec{x}_{+1}| \cos (\omega \eta_{+1}^\prime)} \cos (\omega |\vec{x}_{+1}^\prime|) x_{+1}^{\prime i}\\
Z^4 &= \frac{1}{\omega \cos (\omega \eta_{+1}^\prime)} \sin (\omega |\vec{x}_{+1}^\prime|)
\end{aligned}$ &
 $\begin{aligned}
Z^0 &=\frac{1}{\omega} \tan (\omega \eta_{+1}^\prime) \\
Z^1 &=\frac{1}{\omega \cos (\omega \eta_{+1}^\prime)} \cos (\omega r_{+1}^\prime) \sin \theta \cos \phi \\
Z^2 &=\frac{1}{\omega \cos (\omega \eta_{+1}^\prime)} \cos (\omega r_{+1}^\prime) \sin \theta \sin \phi \\
Z^3 &=\frac{1}{\omega \cos (\omega \eta_{+1}^\prime)} \cos (\omega r_{+1}^\prime) \cos \theta \\
Z^4 &=\frac{1}{\omega \cos (\omega \eta_{+1}^\prime)} \sin (\omega r_{+1}^\prime) 
\end{aligned}$ \\ 

Ranges of & $\eta_{+1}^\prime \in \left[-\frac{\pi}{2\omega},\frac{\pi}{2\omega}\right)$ & $\eta_{+1}^\prime \in \left[-\frac{\pi}{2\omega},\frac{\pi}{2\omega}\right)$ \\
coordinates & $|\vec{x}_{+1}^\prime| \in \left[-\frac{\pi}{2\omega},\frac{\pi}{2\omega}\right)$ & $r_{+1} \in \left[-\frac{\pi}{2\omega},\frac{\pi}{2\omega}\right)$ \\ 
cover & \multicolumn{2}{c}{all the manifold} \\ 

& \multicolumn{2}{c}{conformal comoving} \\

Refs. &  & \cite[rel. (3.18)]{eriksen1995sitter}  \\
  
line elem. &  & $ds^2=\frac{1}{\cos^2 (\omega \eta_{+1}^\prime)}( d\eta_{+1}^{\prime 2}-dr_{+1}^{\prime 2}- \frac{\cos^2(\omega r_{+1}^\prime)}{\omega^2}d\Omega_2^2) $ \\ 

\end{longtable}



\newpage
\begin{longtable}[ht]{ccc}
 \caption{{\bf  k=-1 proper FLRW chart, standard coordinates}}\\ \label{FLRWkmps}
     & cartesian $\{t_{-1},\vec{x}_{-1}\}$ & spherical $\{t_{-1},r_{-1},\theta,\phi\}$ \\ \hline
      
 def. & $\begin{aligned}
Z^0 &=\frac{1}{\omega}\sinh(\omega t_{-1}) \cosh (\omega |\vec{x}_{-1}|) \\
Z^i &= \frac{1}{\omega  |\vec{x}_{+1}|}\sinh(\omega t_{-1}) \sinh (\omega |\vec{x}_{+1}|) x_{-1}^i\\
Z^4 &=\frac{1}{\omega}\cosh(\omega t_{-1}) 
\end{aligned}$ &
 $\begin{aligned}
Z^0 &=\frac{1}{\omega}\sinh(\omega t_{-1}) \cosh (\omega r_{-1}) \\
Z^1 &=\frac{1}{\omega}\sinh(\omega t_{-1}) \sinh (\omega r_{-1}) \sin \theta \cos \phi \\
Z^2 &=\frac{1}{\omega}\sinh(\omega t_{-1}) \sinh (\omega r_{-1}) \sin \theta \sin \phi \\
Z^3 &=\frac{1}{\omega}\sinh(\omega t_{-1}) \sinh (\omega r_{-1}) \cos \theta \\
Z^4 &=\frac{1}{\omega}\cosh(\omega t_{-1}) 
\end{aligned}$ \\ 

Ranges of & $t_{-1} \in (-\infty,+\infty)$ & $t_{-1} \in (-\infty,+\infty)$ \\
coordinates & $|\vec{x}_{-1}| \in \left[0,\infty \right)$ & $r_{-1} \in \left[0,\infty \right)$ \\ 

& \multicolumn{2}{c}{comoving} \\

Refs. &  & \cite[p.50]{griffiths2009exact} \\
  & & \cite[(A6,A7)]{mottola1985particle} \cite[(25,26)]{spradlin2001houches}  \\
  
line elem. &  & $ds^2=dt_{-1}^2-\sinh^2(\omega t_{-1}) (dr_{-1}^2+ \frac{\sinh^2(\omega r_{-1})}{\omega^2}d\Omega_2^2) $ \\ 

$\sqrt{|g|}$& $\frac{\sinh^3 (\omega t_{-1}) \sinh^2 (\omega |\vec{x}_{-1}|)}{\omega^3 \vec{x}_{-1}^2}$ & \\ 
 
\end{longtable}


\begin{longtable}[ht]{ccc}
 \caption{{\bf  k=-1 proper FLRW chart, alternate coordinates}}\\ \label{FLRWkmpa}
     & cartesian $\{t_{-1},\vec{\bar{x}}_{-1}\}$ & spherical $\{t_{-1},\bar{r}_{-1},\theta,\phi\}$ \\ \hline
      
 def. & $\begin{aligned}
Z^0 &=\frac{\sqrt{1+\omega^2 \bar{r}_{-1}^2}}{\omega}\sinh(\omega t_{-1}) \\
Z^i &=\sinh(\omega t_{-1}) \bar{x}^i_{-1} \\
Z^4 &=\frac{1}{\omega}\cosh(\omega t_{-1})
\end{aligned}$ &
 $\begin{aligned}
Z^0 &=\frac{\sqrt{1+\omega^2 \bar{r}_{-1}^2}}{\omega}\sinh(\omega t_{-1}) \\
Z^1 &=\sinh(\omega t_{-1}) \bar{r}_{-1} \sin \theta \cos \phi \\
Z^2 &=\sinh(\omega t_{-1}) \bar{r}_{-1} \sin \theta \sin \phi \\
Z^3 &=\sinh(\omega t_{-1}) \bar{r}_{-1} \cos \theta \\
Z^4 &=\frac{1}{\omega}\cosh(\omega t_{-1})
\end{aligned}$ \\ 

Ranges of & $t_{-1} \in (-\infty,+\infty)$ & $t_{-1} \in (-\infty,+\infty)$ \\
coordinates & $|\vec{\bar{x}}_{-1}| \in \left[0,\infty \right)$ & $\bar{r}_{-1} \in \left[0,\infty \right)$ \\ 

Refs. &  & \cite[rel. (4.3)]{eriksen1995sitter} \\
  
line elem. &  & $ds^2=dt_{-1}^2- \sinh^2(\omega t_{-1}) \left( \frac{d\bar{r}_{-1}^2}{1+\omega^2 \bar{r}_{-1}^2}-\bar{r}_{-1}^2 d\Omega^2_2 \right)$ \\ 

$\sqrt{|g|}$& $ \frac{\sinh^3(\omega t_{-1})}{\sqrt{1+\omega^2 \vec{\bar{x}}^2_{-1}}}$ & \\ 

Killing vectors & $\begin{aligned} 
K_{04}&= \sqrt{1+\omega^2 \vec{x}^2}\left( \frac{1}{\omega}\partial_0+ \right. \\
&+ \left. \coth (\omega t) x^i \partial_i \right) \\
K_{0i}&=\frac{1}{\omega} \sqrt{1+\omega^2 \vec{x}^2} \partial_i \\
K_{i4}&=x_i \partial_0 - \frac{\coth(\omega t)}{\omega}\partial_i+ \\
&+ \omega \coth (\omega t) x_i x^j \partial_j \\

K_{ij}&=x_i\partial_j-x_j\partial_i
\end{aligned}$& \\  

\end{longtable}


\begin{longtable}[ht]{ccc}
 \caption{{\bf k=-1 proper FLRW chart, isotropic coordinates}}\\ \label{FLRWkmpi}
     & cartesian $\{t_{-1},\vec{\bf{x}}_{-1}\}$ & spherical $\{t_{-1},\rho_{-1},\theta,\phi\}$ \\ \hline
      
 def. & $\begin{aligned}
Z^0 &=\frac{1}{\omega}\cosh(\omega t_{-1}) \sqrt{\frac{1+\frac{\omega^2 \bf{x}_{-1}^2}{4}}{1-\frac{\omega^2 \bf{x}_{-1}^2}{4}}} \\
Z^i &=\sinh(\omega t_{-1}) \frac{\bf{x}^i_{-1}}{1-\frac{\omega^2 \bf{x}_{-1}^2}{4}} \\
Z^4 &=\frac{1}{\omega}\cosh(\omega t_{-1})
\end{aligned}$ &
 $\begin{aligned}
Z^0 &=\frac{1}{\omega}\cosh(\omega t_{-1}) \sqrt{1+\frac{\omega^2 \rho_{-1}^2}{1-\frac{\omega^2 \rho_{-1}^2}{4}}} \\
Z^1 &=\sinh(\omega t_{-1}) \frac{\rho_{-1}}{1-\frac{\omega^2 \rho_{-1}^2}{4}} \sin \theta \cos \phi \\
Z^2 &=\sinh(\omega t_{-1}) \frac{\rho_{-1}}{1-\frac{\omega^2 \rho_{-1}^2}{4}}  \sin \theta \sin \phi \\
Z^3 &=\sinh(\omega t_{-1}) \frac{\rho_{-1}}{1-\frac{\omega^2 \rho_{-1}^2}{4}}  \cos \theta \\
Z^4 &=\frac{1}{\omega}\cosh(\omega t_{-1})
\end{aligned}$ \\ 

Ranges of & $t_{-1} \in (-\infty,+\infty)$ & $t_{-1} \in (-\infty,+\infty)$ \\
coordinates & $\vec{\bf{x}}_{-1} \in \mathbb{R}^3$& $\rho_{-1} \in [0,\infty)$ \\ 
& \multicolumn{2}{c}{speed of light is isotropic} \\ 

Refs. &  & \cite[rel. (5.9)]{eriksen1995sitter}\\

line elem. & $ds^2= dt_{-1}^2- \frac{\sinh^2(\omega t_{-1})}{1-\frac{\omega^2 \bf{x}_{-1}^2}{4}} d\vec{\bf{x}}^2$ & $ds^2=dt_{-1}^2- \frac{\sinh^2(\omega t_{-1})}{1-\frac{\omega^2 \rho_{-1}^2}{4}} \left( d\rho_{-1}^2+ \rho_{-1}^2 d\Omega^2_2 \right)$ \\
\end{longtable}


\begin{longtable}[ht]{ccc}
 \caption{{\bf  k=-1 conformal FLRW chart, standard coordinates}}\\ \label{FLRWkmcs}
     & cartesian $\{\eta_{-1},\vec{x}_{-1}\}$ & spherical $\{\eta_{-1},r_{-1},\theta,\phi\}$ \\ \hline
      
 def. & $\begin{aligned}
Z^0 &=-\frac{1}{\omega \sinh(\omega \eta_{-1})} \cosh (\omega |\vec{x}_{-1}|) \\
Z^i &= -\frac{1}{\omega  |\vec{x}_{+1}| \sinh(\omega \eta_{-1})} \sinh (\omega |\vec{x}_{+1}|) x_{-1}^i\\
Z^4 &=-\frac{1}{\omega \tanh(\omega t_{-1})} 
\end{aligned}$ &
 $\begin{aligned}
Z^0 &=-\frac{1}{\omega \cosh(\omega \eta_{-1})} \cosh (\omega r_{-1}) \\
Z^1 &=-\frac{1}{\omega \cosh(\omega \eta_{-1})} \sinh (\omega r_{-1}) \sin \theta \cos \phi \\
Z^2 &=-\frac{1}{\omega \cosh(\omega \eta_{-1})} \sinh (\omega r_{-1}) \sin \theta \sin \phi \\
Z^3 &=-\frac{1}{\omega \cosh(\omega \eta_{-1})} \sinh (\omega r_{-1}) \cos \theta \\
Z^4 &=-\frac{1}{\omega \tanh(\omega \eta_{-1})} 
\end{aligned}$ \\ 

Ranges of & $\eta_{-1} \in (-\infty,0)$ & $\eta_{-1} \in (-\infty,0)$ \\
coordinates & $|\vec{x}_{-1}| \in \left[0,\infty \right)$ & $r_{-1} \in \left[0,\infty \right)$ \\ 

& \multicolumn{2}{c}{comoving} \\

Refs. &  & \cite[rel. (4.9)]{podolsky1997impulsive} \cite[rel. (5.18)]{eriksen1995sitter} \\
  
line elem. &  & $ds^2=\frac{1}{\sinh^2(\omega \eta_{-1})}(d\eta_{-1}^2- dr_{-1}^2-$ \\
& & $-\frac{\sinh^2(\omega r_{-1})}{\omega^2}d\Omega_2^2) $ \\ 

$\sqrt{|g|}$& $\frac{\sinh^2 (\omega |\vec{x}_{-1}|)}{\omega^3 \sinh^3 (\omega \eta_{-1}) \vec{x}_{-1}^2}$ & \\ 
 
\end{longtable}

 \newpage
\begin{longtable}[ht]{cccc}
     \caption{{\bf Overview- standard proper FLRW charts}}\\
        $\times$ & \multicolumn{3}{c}{line element: $ds^2=dt_k^2- a^2(t_k)(dr^2_k + f^2(r_k)d\Omega^2_2)$}   \\ 
        
        spatial curvature & $k=+1$ & $k=0$ & $k=-1$ \\ 
        name & closed & spatially flat & open \\ 
        introduced: & Lanczos, 1922 \cite{lanczos1922bemerkung} & Lem\^{a}itre, 1925 \cite{lemaitre1925note}& Robertson, 1933 \cite{robertson1933relativistic}\\ 
        studied: & & Robertson, 1928 \cite{robertson1928relativistic}& \\
        slices of const $t_k$ & spheres & planes & hyperboloids \\ 
        scale factor $a(t_k)$ & $\cosh(\omega t_{+1})$ & $e^{\omega t}$ & $\sinh(\omega t_{-1})$ \\
\end{longtable}  

 
\begin{longtable}[ht]{cccc}
     \caption{{\bf Overview- standard conformal FLRW charts}}\\
        $\times$ & \multicolumn{3}{c}{line element: $ds^2=\Omega^2(\eta_k)(d\eta_k^2-dr^2_k - f^2(r_k)d\Omega^2_2)$ \cite[p.139]{hawking1973large}}  \\    
        conformal time & $\tan \frac{\omega \eta_{+1}}{2}=e^{\omega t_{+1}}$ & $\eta=-\frac{1}{\omega}e^{-\omega t}$ & $\coth \frac{\omega \eta_{-1}}{2}=-e^{\omega t_{-1}}$ \\ 
        $\eta_k(t_k)=\int \frac{dt_k}{a(t_k)}$ & $\sin (\omega \eta_{+1})= \sech(\omega t_{+1})$ ${}^\dagger$ &  & $e^{\omega \eta_{-1}}=\tanh\frac{\omega t_{-1}}{2}$\\ 
        conformal & $ds^2=\frac{1}{\sin^2 \eta_{+1}}(d\eta_{+1}^2-$ & $ds^2=\frac{1}{\omega^2 \eta^2}( d\eta^2-$ & $ds^2=\frac{1}{\sinh^2 \eta_{-1}}(d\eta_{-1}^2-$ \\ 
        line element & $-dr_{+1}^2-\frac{\sin^2 r_{+1}}{\omega^2}d\Omega^2_2)$ & $-dr^2-r^2 d\Omega^2_2)$ & $-dr_{-1}^2-\frac{\sinh^2 r_{-1}}{\omega^2}d\Omega^2_2)$ \\
        chart is & part of Einstein & part of Minkowski & part of Open \\ 
        conformal to & Static Universe & spacetime & Einstein Universe \\ 
        with conf. factor & $\frac{1}{\omega \sin \eta_{+1}}$ & $\frac{1}{\omega \eta_0}$ & $\frac{1}{\omega \sinh \eta_{-1}}$ \\
        range & $\eta_{+1} \in \left(0,\frac{\pi}{\omega} \right)$ & $\eta \in (-\infty,0)$ & $\eta_{-1} \in (-\infty,0)$ \\ 
		cover & entire manifold & half of the manifold & \\
        extended range & $-$ & $\eta \in (-\infty,+\infty)$ &  \\ 
        cover (not smoothly) & $-$ & entire manifold & \\ 	                     
\end{longtable}  


 
\newsection{Static charts}

\begin{longtable}[h]{c|c}
   \caption{{\bf Static radial coordinates on de Sitter spacetime}} \\
           $r_s$ & $r_s=\frac{\bar{r_s}}{\sqrt{1+\omega^2 \bar{r}_s^2}}=\frac{1}{\omega} \sin (\omega r_F)= \frac{1}{\omega} \tanh (\omega r^*)= \frac{\rho_s}{1+\frac{\omega^2 \rho_s^2}{4}}$  \\ 
            & $dr_s=\frac{d\bar{r_s}}{(\sqrt{1+\omega^2 \bar{r}_s^2})^\frac{3}{2}}=\cos (\omega r_F) dr_F= \frac{dr^*}{ \cosh (\omega r^*)}= 4\frac{1-\frac{\omega^2 \rho_s^2}{4}}{1+\frac{\omega^2 \rho_s^2}{4}}d\rho_s$  \\ 					
						$\bar{r}_s$ & $\bar{r}_s=\frac{r_s}{\sqrt{1-\omega^2 r_s^2}}=\frac{1}{\omega} \tan (\omega r_F)= \frac{1}{\omega} \sinh (\omega r^*)= \frac{\rho_s}{1-\frac{\omega^2 \rho_s^2}{4}}$  \\ 
            & $d\bar{r}_s=\frac{dr_s}{(\sqrt{1-\omega^2 r_s^2})^\frac{3}{2}}=\frac{dr_F}{\cos (\omega r_F)} = \cosh (\omega r^*) dr^* = -4\frac{1+\frac{\omega^2 \rho_s^2}{4}}{1-\frac{\omega^2 \rho_s^2}{4}}d\rho_s$  \\ 					
           $r_F$ & $r_F=\frac{1}{\omega} \arcsin (\omega r_s)=\frac{1}{\omega} \arcsin \frac{\omega \bar{r}_s}{\sqrt{1+\omega^2 \bar{r}_s^2}}= \frac{1}{\omega} \arcsin (\tanh (\omega r^*))= \frac{1}{\omega} \arcsin \frac{\omega \rho_s}{1+\frac{\omega^2 \rho_s^2}{4}}$  \\ 
           & $dr_F=\frac{dr_s }{\sqrt{1-\omega^2 r_s^2}}=\frac{d\bar{r}_s }{1+\omega^2 \bar{r}_s^2}= \frac{dr^*}{\cosh (\omega r^*)}= -  \frac{d\rho_s}{1+\frac{\omega^2 \rho_s^2}{4}}$  \\ 
           $r^*$ & $r^*=\frac{1}{\omega} \arctanh (\omega r_s)=\frac{1}{\omega} \arctanh \frac{\omega \bar{r}_s}{\sqrt{1+\omega^2 \bar{r}_s^2}}= \frac{1}{\omega} \arctanh (\sin (\omega r^*))= \frac{1}{\omega} \arctanh \frac{\omega \rho_s}{1+\frac{\omega^2 \rho_s^2}{4}}$  \\
\end{longtable}

\begin{longtable}[ht]{ccc}
 \caption{{\bf Static chart}}\\ \label{static}
 Introduced:& \multicolumn{2}{c}{Eddington, 1922 \cite{eddington1922mathematical}}\\
     & cartesian $\{t_s,\vec{x}_s\}$ & spherical $\{t_s,r_s,\theta,\phi\}$ \\ \hline
      
 def. & $\begin{aligned}
Z^0 &=\frac{\sqrt{1-\omega^2 \vec{x}_s^2}}{\omega}\sinh (\omega t_s) \\
Z^i &= x_s^i \\
Z^4 &=\frac{\sqrt{1-\omega^2 \vec{x}_s^2}}{\omega}\cosh (\omega t_s) 
\end{aligned}$ &
 $\begin{aligned}
Z^0 &=\frac{\sqrt{1-\omega^2 r_s^2}}{\omega}\sinh (\omega t_s) \\
Z^1 &=r_s \sin \theta \cos \phi \\
Z^2 &=r_s \sin \theta \sin \phi \\
Z^3 &=r_s \cos \theta \\
Z^4 &=\frac{\sqrt{1-\omega^2 r_s^2}}{\omega}\cosh (\omega t_s) 
\end{aligned}$ \\ 

Ranges of & $t_s \in (-\infty,+\infty)$ & $t_s \in (-\infty,+\infty)$ \\
coordinates & $|\vec{x}_s| \in \left[0,\frac{1}{\omega}\right)$ & $r_s \in \left[0,\frac{1}{\omega}\right)$ \\ 
cover & \multicolumn{2}{c}{region $Z^0+Z^1>0$ (half of the manifold)} \\ 
& \multicolumn{2}{c}{has coordinate singularity at $|\vec{x}_s|=r_s=\frac{1}{\omega}$} \\
\end{longtable}

 
\begin{longtable}[h]{ccc}
   \caption{{\bf Alternate de Sitter static chart}}\\ \label{statica}
 Introduced:& \multicolumn{2}{c}{de Sitter, 1917 \cite{desitter1917einstein} \cite{desitter2_hist}}\\   
     & cartesian $\{t_s,\vec{\bar{x}}_s\}$ & spherical $\{t_s,\bar{r}_s,\theta,\phi\}$ \\ \hline
       
 def. & $\begin{aligned}
Z^0 &=\frac{1}{\omega\sqrt{1+\omega^2 \vec{\bar{x}}_s^2}}\cosh (\omega t_s) \\
Z^i &= \frac{\bar{x}_s^i}{\sqrt{1+\omega^2 \vec{\bar{x}}_s^2}} \\
Z^4 &=\frac{1}{\omega\sqrt{1+\omega^2 \vec{\bar{x}}_s^2}}\sinh (\omega t_s)
\end{aligned}$ &
 $\begin{aligned}
Z^0 &=\frac{1}{\omega\sqrt{1+\omega^2 \vec{\bar{r}}_s^2}}\cosh (\omega t_s) \\
Z^1 &=\frac{\bar{r}_s}{\sqrt{1+\omega^2 \vec{\bar{r}}_s^2}} \sin \theta \cos \phi \\
Z^2 &=\frac{\bar{r}_s}{\sqrt{1+\omega^2 \vec{\bar{r}}_s^2}} \sin \theta \sin \phi \\
Z^3 &=\frac{\bar{r}_s}{\sqrt{1+\omega^2 \vec{\bar{r}}_s^2}} \cos \theta \\
Z^4 &=\frac{1}{\omega\sqrt{1+\omega^2 \vec{\bar{r}}_s^2}}\sinh (\omega t_s)
\end{aligned}$ \\       

Ranges of & $t_s \in (-\infty,+\infty)$ & $t_s \in (-\infty,+\infty)$ \\
coordinates & $|\vec{\bar{x}}_s| \in \left[0,\frac{1}{\omega}\right)$ & $\bar{r}_s \in [0,\frac{1}{\omega})$ \\
covers & \multicolumn{2}{c}{(same as the static chart)} \\
Refs. &  &  \cite[rel. (2.7)]{schmidt1993sitter} \\
		
metric &   & $ds^2=\frac{1}{1+\omega^2 \bar{r}_s^2}(dt^2-$ \\ 
 &  & $- \frac{d\bar{r}^2}{1+\omega^2 \bar{r}_s^2} -\bar{r}_s^2 d\Omega^2_2) $ \\ 
$\sqrt{|g|}$& $ \frac{\sqrt{1-\omega^2 \vec{\bar{x}}^2}}{(1+\omega^2 \vec{\bar{x}}^2)^3}$  &  \\ 
\end{longtable}

\begin{longtable}[h]{ccc}
   \caption{{\bf Fermi (original de Sitter static) chart}}\\ \label{staticf}
 Introduced:& \multicolumn{2}{c}{de Sitter, 1917 \cite{desitter1917einstein} \cite{desitter2_hist}}\\   
     & cartesian $\{t_s,\vec{x}_F\}$ & spherical $\{t_s,r_F,\theta,\phi\}$ \\ \hline
       
	 def. & $\begin{aligned}
Z^0 &=\frac{\cos(\omega |\vec{x}_F|)}{\omega}\sinh (\omega t_s) \\
Z^i &= \frac{\sin(\omega |\vec{x}_F|)}{\omega |\vec{x}_F|} x_F^i \\
Z^4 &=\frac{\cos(\omega |\vec{x}_F|)}{\omega}\cosh (\omega t_s)
\end{aligned}$ &
 $\begin{aligned}
Z^0 &=\frac{\cos(\omega r_F)}{\omega}\sinh (\omega t_s) \\
Z^1 &=\frac{\sin(\omega r_F)}{\omega} \sin \theta \cos \phi \\
Z^2 &=\frac{\sin(\omega r_F)}{\omega} \sin \theta \sin \phi \\
Z^3 &=\frac{\sin(\omega r_F)}{\omega} \cos \theta \\
Z^4 &=\frac{\cos(\omega r_F)}{\omega}\cosh (\omega t_s)
\end{aligned}$ \\ 		

Ranges of & $t_s \in (-\infty,+\infty)$ & $t_s \in (-\infty,+\infty)$ \\
coordinates & $|\vec{x}_F| \in \left[0,\frac{\pi}{2\omega} \right) $ & $r_F \in \left[0,\frac{\pi}{2\omega} \right)$ \\
covers & \multicolumn{2}{c}{(same as the static chart)} \\
Refs. & \cite[rel. (25,28,29)]{chicone2006explicit} &  \cite[rel. (31)]{chicone2006explicit} \cite[rel. (1.1)]{eriksen1995sitter} \cite[rel. (2.5)]{schmidt1993sitter} \\
			
metric & $ds^2= \cos^2(\omega |\vec{x}_F|) - \left( \frac{x_{Fi}x_{Fj}}{\vec{x}_F^2}+ \right.$ & $ds^2=\cos^2(\omega r_F)dt_s^2- dr_F^2$ \\ 
 & $+ \left. \frac{\sin^2(\omega |\vec{x}_F|)}{\omega^2 |\vec{x}_F|}(\delta_{ij}- \frac{x_{Fi}x_{Fj}}{\vec{x}_F^2}) \right) dx_F^i dx_F^j$ & $- \frac{\sin^2 (\omega r_F)}{\omega^2}  d\Omega^2_2 $ \\ 
$\sqrt{|g|}$& $ \frac{\sin^2 (\omega |\vec{\bar{x}}|) \cos (\omega |\vec{\bar{x}}|)}{\omega^2 \vec{\bar{x}}^2}$  &  \\ 
\end{longtable}

\begin{longtable}[h]{ccc}
   \caption{{\bf Tortoise (Regge- Wheeler static) chart}}\\ \label{staticr}
     & cartesian $\{t_s,\vec{x}^*\}$ & spherical $\{t_s,r^*,\theta,\phi\}$ \\ \hline
 
	 def. & $\begin{aligned}
Z^0 &=\frac{1}{\omega \cosh (\omega |\vec{x}^*|)}\cosh (\omega t_s) \\
Z^i &= \frac{\tanh(\omega |\vec{x}^*|)}{\omega |\vec{x}^*|} x^{*i} \\
Z^4 &=\frac{1}{\omega \cosh (\omega |\vec{x}^*|)}{\omega}\sinh (\omega t_s)
\end{aligned}$ &
 $\begin{aligned}
Z^0 &=\frac{1}{\omega \cosh (\omega r^*)}\cosh (\omega t_s) \\
Z^1 &=\frac{\tanh(\omega r^*)}{\omega r^*} \sin \theta \cos \phi \\
Z^2 &=\frac{\tanh(\omega r^*)}{\omega r^*} \sin \theta \sin \phi \\
Z^3 &=\frac{\tanh(\omega r^*)}{\omega r^*} \cos \theta \\
Z^4 &=\frac{1}{\omega \cosh (\omega r^*)}\sinh (\omega t_s)
\end{aligned}$ \\ 	

Ranges of & $t_s \in (-\infty,+\infty)$ & $t_s \in (-\infty,+\infty)$ \\
coordinates & $\vec{x}^* \in \mathbb{R}^3_- $ & $r^* \in (-\infty,0)$ \\
covers & \multicolumn{2}{c}{(same as the static chart)} \\
Refs. & &  \cite[rel. (18)]{beck1984kruskal} \\

metric &  & $ds^2=\frac{1}{\cosh^2 (\omega r^*)}(dt_s^2-dr^{*2}-$ \\ 
 &  & $- \frac{\sinh^2 (\omega r^*)}{\omega^2}  d\Omega^2_2)$ \\ 
\end{longtable}

\begin{longtable}[ht]{ccc}
 \caption{{\bf Static chart, isotropic coordinates}}\\ \label{statici}
     & cartesian $\{t_s,\vec{\bf{x}}_s\}$ & spherical $\{t_s,\rho_s,\theta,\phi\}$ \\ \hline
      
 def. & $\begin{aligned}
Z^0 &=\frac{1-\frac{\omega^2 \vec{\bf{x}}^2_s}{4}}{1+\frac{\omega^2 \vec{\bf{x}}^2_s}{4}}  \sinh (\omega t_s) \\
Z^i &= \frac{\vec{\bf{x}}^i_s}{1+\frac{\omega^2 \vec{\bf{x}}^2_s}{4}} \\
Z^4 &= \frac{1-\frac{\omega^2 \vec{\bf{x}}^2_s}{4}}{1+\frac{\omega^2 \vec{\bf{x}}^2_s}{4}} \cosh (\omega t_s) 
\end{aligned}$ &
 $\begin{aligned}
Z^0 &=\frac{1-\frac{\omega^2 \rho_s^2}{4}}{1+\frac{\omega^2 \rho_s^2}{4}} \sinh (\omega t_s) \\
Z^1 &=\frac{\rho_s}{1+\frac{\omega^2 \rho_s^2}{4}} \sin \theta \cos \phi \\
Z^2 &=\frac{\rho_s}{1+\frac{\omega^2 \rho_s^2}{4}} \sin \theta \sin \phi \\
Z^3 &=\frac{\rho_s}{1+\frac{\omega^2 \rho_s^2}{4}} \cos \theta \\
Z^4 &=\frac{1-\frac{\omega^2 \rho_s^2}{4}}{1+\frac{\omega^2 \rho_s^2}{4}} \cosh (\omega t_s) 
\end{aligned}$ \\ 

Ranges of & $t_s \in (-\infty,+\infty)$ & $t_s \in (-\infty,+\infty)$ \\
coordinates & $|\vec{\bf{x}}_s| \in \left[0,\frac{2}{\omega}\right)$ & $\rho_s \in \left[0,\frac{2}{\omega}\right)$ \\ 
covers & \multicolumn{2}{c}{(same as the static chart)} \\

Refs. &  & \cite[rel. (5.10-11)]{eriksen1995sitter}  \\
  
line elem. &  & $ds^2= \frac{1}{\left(1+\frac{\omega^2 \rho_s^2}{4} \right)^2} \left [ \left( 1-\frac{\omega^2 \rho_s^2}{4} \right)^2 dt_s^2- \right.$ \\ 
 &  & $ \left. -d\rho_s^2- \rho_s^2 d\Omega^2_2 \right]$ \\ 
\end{longtable}

\newsection{'Natural' charts}
\begin{longtable}[ht]{ccc}
  \caption{{\bf Minkowskian chart}}\\
     & cartesian $\{t_M,\vec{x}_s\}$ & spherical $\{t_M,r_s,\theta,\phi\}$ \\ \hline
 def. & $\begin{aligned}
Z^0 &=t_M \\
Z^i &=x_s^i \\
Z^4 &=\frac{1}{\omega} \sqrt{1+\omega^2(t_M^2-\vec{x}_s^2)} \\
\end{aligned}$ &
 $\begin{aligned}
Z^0 &=t_M \\
Z^1 &=r_s \sin \theta \cos \phi \\
Z^2 &=r_s \sin \theta \sin \phi \\
Z^3 &=r_s \cos \theta \\
Z^4 &=\frac{1}{\omega} \sqrt{1+\omega^2(t_M^2-r_s^2)} \\
\end{aligned}$ \\ 
cover & \multicolumn{2}{c}{half of the manifold \cite{candelas1975general}} \\

& \multicolumn{2}{c}{} \\ 
Refs. & \cite[sec. 2.2]{mignemi2008doubly} & \\

line elem. &   & $ds^2=\frac{1}{1+\omega^2(t_M^2-r_s^2)}\Big( (1-\omega^2 r_s^2)dt_M^2+$ \\ 
 &   & $+2\omega^2 t_M r_s dt_M dr_s-(1+\omega t_M^2)dr_s^2 \Big) -r_s^2 d\Omega^2_2$ \\ 
& $ g_{\mu\nu}=\eta_{\mu\nu}-\frac{\omega^2 x_{s\mu}x_{s\nu}}{1+\omega^2(t_M^2-\vec{x}_s^2)}$& \\
& $ g^{\mu\nu}=\eta^{\mu\nu}+\omega^2 x_s^\mu x_s^\nu$& \\

$\sqrt{|g|}$& $ (1-\omega^2(t_M^2-\vec{x}_s^2))^{-\frac{1}{2}} $ & \\ 

a tetrad & $\left\{ \begin{aligned}
 \hat{e}^{\hat{\alpha}}_{\mu}&= \delta^\alpha_\mu +  \frac{1}{t_M^2-\vec{x}_s^2} \cdot \\  \cdot & \left(\sqrt{1+ \omega^2(t_M^2-\vec{x}_s^2)} -1 \right)\eta_{\mu\theta}x^\theta_M x^\alpha_M
\end{aligned}\right.$ & \\ 
& $\left\{ \begin{aligned}
 e_{\hat{\alpha}}^{\mu}&= \delta_\alpha^\mu +  \frac{1}{t_M^2-\vec{x}_s^2} \cdot \\  \cdot & \left(\frac{1}{\sqrt{1+ \omega^2(t_M^2-\vec{x}_s^2)}} -1 \right)\eta_{\alpha\theta}x^\theta_M x^\mu_M
\end{aligned}\right. $ & \\ \\

Killing vectors & $\begin{aligned} 
K_{\mu 4}&=\frac{1}{\omega}  \sqrt{1 +\omega^2 (t^2-\vec{x}^2)} \partial_\mu \\
K_{\mu \nu}&=x_\mu \partial_\nu-x_\nu\partial_\mu
\end{aligned}$& \\  
\end{longtable}

\newpage
\begin{longtable}[ht]{ccc}
  \caption{{\bf Beltrami chart}}\\
     & cartesian $\{t_B,\vec{x}_B\}$ & spherical $\{t_B,r_B,\theta,\phi\}$ \\ \hline
 def. & $\begin{aligned}
Z^0 &=\frac{t_B}{\sqrt{1-\omega^2(t_B^2-\vec{x}_B^2)}} \\
Z^i &=\frac{x_B^i}{\sqrt{1-\omega^2(t_B^2-\vec{x}_B^2)}} \\
Z^4 &=\frac{1}{\omega \sqrt{1-\omega^2(t_B^2-\vec{x}_B^2)}} \\
\end{aligned}$ &
 $\begin{aligned}
Z^0 &=\frac{t_B}{\sqrt{1-\omega^2(t_B^2-r_B^2)}} \\
Z^1 &=\frac{r_B \sin \theta \cos \phi}{\sqrt{1-\omega^2(t_B^2-r_B^2)}} \\
Z^2 &=\frac{r_B \sin \theta \sin \phi}{\sqrt{1-\omega^2(t_B^2-r_B^2)}} \\
Z^3 &=\frac{r_B \cos \theta}{\sqrt{1-\omega^2(t_B^2-r_B^2)}} \\
Z^4 &=\frac{1}{\omega \sqrt{1-\omega^2(t_B^2-r_B^2)}} \\
\end{aligned}$ \\

cover & \multicolumn{2}{c}{1/8 of the manifold \cite{guo2003beltrami}} \\
& \multicolumn{2}{c}{} \\ 
Refs. & \cite{mallett1973position} \cite{guo2003beltrami} \cite{wang2008classic} \cite[sec. 2.4]{mignemi2008doubly}  & \\

& $ g_{\mu\nu}=\frac{\left(1-\omega^2(t_B^2-\vec{x}_B^2)\right)\eta_{\mu\nu}+\omega^2 x_{B\mu}x_{B\nu}}{\left(1-\omega^2(t_B^2-\vec{x}_B^2)\right)^2}$ & \\

& $ g^{\mu\nu}=\left(1-\omega^2(t_B^2-\vec{x}_B^2)\right)(\eta^{\mu\nu}-\omega^2 x_B^\mu x_B^\nu)$ & \\
\end{longtable}

\newpage
\begin{longtable}[ht]{ccc}
  \caption{{\bf Stereographic (conformally Minkowski) chart}}\\
     & cartesian $\{t_{st},\vec{x}_{st}\}$ & spherical $\{t,r_{st},\theta,\phi\}$ \\ \hline
 def. & $\begin{aligned}
Z^0 &=\frac{t_{st}}{1-\frac{\omega^2(t_{st}^2-\vec{x}_{st}^2)}{4}} \\
Z^i &=\frac{x_{st}^i}{1-\frac{\omega^2(t_{st}^2-\vec{x}_{st}^2)}{4}} \\
Z^4 &= \frac{1}{\omega}\left(1-\frac{2}{1-\frac{\omega^2(t_{st}^2-\vec{x}_{st}^2)}{4}}\right) \\
\end{aligned}$ &
 $\begin{aligned}
Z^0 &=\frac{t_{st}}{1-\frac{\omega^2(t_{st}^2-\vec{x}_{st}^2)}{4}} \\
Z^1 &=\frac{r_{st}\sin \theta \cos \phi}{1-\frac{\omega^2(t_{st}^2-r_{st}^2)}{4}} \\
Z^2 &=\frac{r_{st}\sin \theta \sin \phi}{1-\frac{\omega^2(t_{st}^2-r_{st}^2)}{4}} \\
Z^3 &=\frac{r_{st}\cos \theta}{1-\frac{\omega^2(t_{st}^2-r_{st}^2)}{4}} \\
Z^4 &= \frac{1}{\omega}\left(1-\frac{2}{1-\frac{\omega^2(t_{st}^2-r_{st}^2)}{4}}\right) \\
\end{aligned}$ \\ 

cover & \multicolumn{2}{c}{3/4 of the manifold}  \\
& \multicolumn{2}{c}{has cosmological horizon at $|\vec{x}_{st}|=r_{st}=\frac{2}{\omega}$} \\ 

Refs. & \cite[sec. 2.3]{mignemi2008doubly} \cite[sec. A4]{bicak2005fields}& \\
& \cite{aldrovandi2007sitter} \cite{aldrovandi2009sitter} & \\

& $ g_{\mu\nu}=\frac{\eta_{\mu\nu}}{1-\frac{\omega^2(t_{st}^2-\vec{x}_{st}^2)}{4}}$ & \\
\end{longtable}

\newsection{'Hybrid' charts}

\begin{longtable}[ht]{ccc}
  \caption{{\bf de Sitter-Painlev\'{e} chart}}\\
     & cartesian $\{t,\vec{x}_s\}$ & spherical $\{t,r_s,\theta,\phi\}$ \\ \hline
 def. & $\begin{aligned}
Z^0 &=\frac{1}{\omega}\sinh (\omega t)+\frac{\omega \vec{x}_s^2 e^{-\omega t}}{2} \\
Z^i &= x_{si} \\
Z^4 &=\frac{1}{\omega}\cosh (\omega t)-\frac{\omega \vec{x}_s^2 e^{-\omega t}}{2}
\end{aligned}$ &
 $\begin{aligned}
Z^0 &=\frac{1}{\omega}\sinh (\omega t)+\frac{\omega r_s^2 e^{-\omega t}}{2} \\
Z^1 &=r_s \sin \theta \cos \phi \\
Z^2 &=r_s \sin \theta \sin \phi \\
Z^3 &=r_s \cos \theta \\
Z^4 &=\frac{1}{\omega}\cosh (\omega t)-\frac{\omega r_s^2 e^{-\omega t}}{2}
\end{aligned}$ \\      
 
cover & \multicolumn{2}{c}{half of the manifold (same as spatially flat FLRW)} \\ 
& \multicolumn{2}{c}{constant time slices are euclidean spaces} \\ 
& \multicolumn{2}{c}{generator of $t$ is a Killing vector} \\ 

Refs. &  & \cite[rel. (2.3)]{gautreau1983geodesic} \\
  & &  \cite[rel. (4)]{parikh2002coordinates} \\
\end{longtable}

\newpage
\newsection{Anisotropic charts}


\begin{longtable}[ht]{cc}
  \caption{{\bf Coordinates for Kantowski-Sachs and Bianchi III charts}}\\
	
$t_s=r_{KS}$ & $t_s=\frac{1}{2\omega} \ln \left| \frac{\sinh (\omega r_{KS}) +\tanh (\omega t_{B3}) \cosh \Theta}{\sinh (\omega r_{KS}) - \tanh (\omega t_{B3}) \cosh \Theta}\right|$ \\
$\sinh (\omega t_{KS})=\frac{1}{\sinh (\omega \bar{t}_{KS})}$ & $\tanh (\omega t_{B3})= \sin (\omega \bar{t}_{B3})$ \\
$r_s=\frac{1}{\omega} \cosh (\omega t_{KS})= \frac{1}{\omega} \coth (\omega t_{KS})$ & $r_s=\sqrt{\sinh^2 (\omega t_{B3}) \sinh^2 \Theta+ \cosh^2(\omega t_{B3}) \cos (\omega r_{B3})}$ \\
& $\tan \theta=\frac{\sinh (\omega t_{B3}) \sinh \Theta}{\cos (\omega r_{B3})}$ \\
\multicolumn{2}{c}{\cite[p. 150]{eriksen1995sitter}} \\
\end{longtable}

\begin{longtable}[ht]{ccc}
  \caption{{\bf Kantowski-Sachs charts}}\\
     & $\{t_{KS},r_{KS},\theta,\phi\}$ 
		& alternate $\{\bar{t}_{KS},r_{KS},\theta,\phi\}$ \\ \hline
		
 def. & $\begin{aligned}
Z^0 &=\frac{1}{\omega}\sinh (\omega t_{KS}) \cosh (\omega r_{KS}) \\
Z^1 &=\frac{1}{\omega}\cosh (\omega t_{KS}) \sin \theta \cos \phi \\
Z^2 &=\frac{1}{\omega}\cosh (\omega t_{KS}) \sin \theta \sin \phi \\
Z^3 &=\frac{1}{\omega}\cosh (\omega t_{KS}) \cos \theta \\
Z^4 &=\frac{1}{\omega}\sinh (\omega t_{KS}) \sinh (\omega r_{KS})
\end{aligned}$ &
 $\begin{aligned}
Z^0 &=\frac{1}{\omega\sinh (\omega \bar{t}_{KS})} \cosh (\omega r_{KS}) \\
Z^1 &=\frac{1}{\omega}\coth (\omega \bar{t}_{KS}) \sin \theta \cos \phi \\
Z^2 &=\frac{1}{\omega}\coth (\omega \bar{t}_{KS}) \sin \theta \sin \phi \\
Z^3 &=\frac{1}{\omega}\coth (\omega \bar{t}_{KS}) \cos \theta \\
Z^4 &=\frac{1}{\omega\sinh (\omega \bar{t}_{KS})} \sinh (\omega r_{KS})
\end{aligned}$ \\ 

cover & \multicolumn{2}{c}{describe the part of the static de Sitter chart on the other }  \\
& \multicolumn{2}{c}{side of the cosmological horizon at $r_{KS}=\frac{1}{\omega}$} \\ 

Refs. & \cite[p. 149]{eriksen1995sitter} \cite[p. 53]{griffiths2009exact} \cite{gron1986transition} \cite{torrence1988note} & \cite[p. 149]{eriksen1995sitter}  \\

line elem. &  $ds^2=dt_{KS}^2-\sinh^2 (\omega t_{KS}) dr_{KS}^2-$ & $ds^2=\frac{1}{\sinh^2(\omega \bar{t}_{KS})} (d\bar{t}_{KS}^2+$ \\ 
 &  $-\frac{\cosh^2(\omega t_{KS})}{\omega^2}d\Omega_2^2$ & $+dr_{KS}^2-\frac{\cosh^2(\omega \bar{t}_{KS})}{\omega^2}d\Omega_2^2)$ \\ 

$\sqrt{|g|}$ & $ \frac{\cosh^2 (\omega t_{KS}) \sinh (\omega t_{KS}) \sin \theta}{\omega^3}$ & $ \frac{\cosh^2 (\omega \bar{t}_{KS})  \sin \theta}{\omega^3 \sinh^3 (\omega \bar{t}_{KS})}$ \\ 

\end{longtable}

\newpage
\begin{longtable}[ht]{ccc}
  \caption{{\bf Bianchi III dS charts}}\\
     &  $\{t_{B3},r_{B3},\Theta,\phi\}$ & alternate $\{\bar{t}_{B3},r_{B3},\Theta,\phi\}$ \\ \hline
		
 def. &  $\begin{aligned}
Z^0 &=\frac{1}{\omega}\sinh (\omega t_{B3}) \cosh \Theta \\
Z^1 &=\frac{1}{\omega}\sinh (\omega t_{B3}) \sinh \Theta \cos \phi \\
Z^2 &=\frac{1}{\omega}\sinh (\omega t_{B3}) \sinh \Theta \sin \phi \\
Z^3 &=\frac{1}{\omega}\cosh (\omega t_{B3}) \cos (\omega r_{B3}) \\
Z^4 &=\frac{1}{\omega}\cosh (\omega t_{B3}) \sin (\omega r_{B3})
\end{aligned}$ &
 $\begin{aligned}
Z^0 &=\frac{1}{\omega}\tan (\omega \bar{t}_{B3}) \cosh \Theta \\
Z^1 &=\frac{1}{\omega}\tan (\omega \bar{t}_{B3}) \sinh \Theta \cos \phi \\
Z^2 &=\frac{1}{\omega}\tan (\omega \bar{t}_{B3}) \sinh \Theta \sin \phi \\
Z^3 &=\frac{1}{\omega \cos (\omega \bar{t}_{B3})} \cos (\omega r_{B3}) \\
Z^4 &=\frac{1}{\omega \cos (\omega \bar{t}_{B3})} \sin (\omega r_{B3})
\end{aligned}$ \\ 

Refs. &  \cite[p. 150]{eriksen1995sitter} \cite[p. 53]{griffiths2009exact} \cite{crawford1993bianchi} &\cite[p. 150]{eriksen1995sitter} \\

line elem. &  $ds^2=dt_{B3}^2-\cosh^2 (\omega t_{B3}) dr_{B3}^2-$ & $ds^2=\frac{1}{\cos(\omega \bar{t}_{B3})}(d\bar{t}_{B3}^2-$ \\ 
 & $-\frac{\sinh^2(\omega t_{B3})}{\omega^2}dH_2^2$  & $-dr_{B3}^2- \frac{\sin^2 (\omega \bar{t}_{B3})}{\omega^2} dH^2_2)$ \\ 

$\sqrt{|g|}$& $ \frac{\cosh (\omega t_{B3}) \sinh^2 (\omega t_{B3}) \sinh \Theta}{\omega^3}$ & $ \frac{\sin^2 (\omega \bar{t}_{B3})  \sinh \Theta}{\omega^3 \cos^3 (\omega \bar{t}_{B3})}$ \\ 
\end{longtable}

\newpage
\begin{longtable}[ht]{cc}
  \caption{{\bf Coordinates for conformally Rindler dS chart}}\\
	
	$ \xi=\frac{\sqrt{1-\omega^2 r_s^2}}{\omega (1-\omega r_s \cos \theta)} $ &  $r=\frac{1}{\omega} \frac{1+\omega^2(- \xi^2+y^2+z^2)}{1+\omega^2(\xi^2+y^2+z^2)}$ \\
	$ y=\frac{r_s}{1-\omega r_s \cos \theta} \sin \theta \cos \phi$ & $\theta=\arctan \frac{2\omega \sqrt{y^2+z^2}}{1-\omega^2(\xi^2+y^2+z^2)}$\\
	$ z=\frac{r_s}{1-\omega r_s \cos \theta} \sin \theta \sin \phi$ & $\phi=\arctan \frac{z}{y}$\\
	& \\
	$\sin \theta= \frac{2\omega \sqrt{y^2+z^2}}{1+\omega^2(-\xi^2+y^2+z^2)}$ & $\cos \theta=\frac{1-\omega^2(\xi^2+y^2+z^2)}{1+\omega^2(-\xi^2+y^2+z^2)}$ \\
	$\sin \phi=\frac{z}{\sqrt{y^2+z^2}}$ & 	$\cos \phi=\frac{y}{\sqrt{y^2+z^2}}$
\end{longtable}

\begin{longtable}[ht]{cc}
  \caption{{\bf Conformally Rindler dS chart}}\\
     & cartesian $\{t_s,\xi,y,z\}$ \\ \hline
		
 def. & 
 $\begin{aligned}
Z^0 &=\frac{2\xi}{1+\omega^2(\xi^2+y^2+z^2)}\sinh (\omega t_s) \\
Z^1 &=\frac{2y}{1+\omega^2(\xi^2+y^2+z^2)} \\
Z^2 &=\frac{2z}{1+\omega^2(\xi^2+y^2+z^2)} \\
Z^3 &=\frac{1-\omega^2(\xi^2+y^2+z^2)}{1+\omega^2(\xi^2+y^2+z^2)} \\
Z^4 &=\frac{2\xi}{1+\omega^2(\xi^2+y^2+z^2)} \cosh (\omega t_s) 
\end{aligned}$ \\ 


Refs. & \cite[p. 187]{birrell1984quantum} \cite{saharian2004casimir} \\

line elem. & $ds^2=\frac{4}{1+\omega^2(\xi^2+y^2+z^2)}(\omega^2 \xi^2 dt_s^2 -d\xi^2 -dy^2-dz^2)$ \\ 
$\sqrt{|g|}$& $ \frac{16\omega \xi}{(1+\omega^2(\xi^2+y^2+z^2))^4}$  \\ 

\end{longtable}

\newpage
\newsection{Charts with null coordinates}
\begin{longtable}[ht]{cccc} 
\caption{{\bf Other coordinates on de Sitter spacetime}} \\   
        Coordinate & Symbol & Relations & References \\ \hline
        Static time & $t_s$  & $t_s=\frac{u +v}{2}$ & - \\     
         & \cite[$t$]{beck1984kruskal,spradlin2001houches} \cite[$T$]{eriksen1995sitter} & $t_s=\frac{1}{\omega}\arctanh \frac{\tilde{V}}{\tilde{U}}$ & \cite[from (7)]{beck1984kruskal}\\
        
         Static coordinate   &  $r_s$ & $r_s=\frac{1}{\omega} \frac{1-\omega^2(\tilde{U}^2+\tilde{V}^2)}{1+\omega^2(\tilde{U}^2-\tilde{V}^2)}$ & \cite[from (6)]{beck1984kruskal} \\
         &  & $r_s= \frac{1}{\omega}\frac{1+\omega^2 UV}{1- \omega^2 UV}$ & \cite[rel. (23)]{spradlin2001houches}\\   

        Tortoise coordinate & $r^*$ &  $r^*=\frac{v -u}{2}$ & - \\ 
                                
        Outgoing (retarded) null  & $u$  & $u=t_s-r^*$ & \cite[rel. (19)]{beck1984kruskal}  \\
\end{longtable}

\begin{longtable}[h]{ccc}
   \caption{{\bf Eddington-Finkelstein charts}}\\
     & ingoing $\{v,r_s,\theta,\phi\}$ & outgoing $\{u,r_s,\theta,\phi\}$ \\ \hline
       
			 def. & $\begin{aligned}
Z^0 &=\frac{1}{\omega} \sinh (\omega v)- r_s \cosh (\omega r_s) \\
Z^1 &=r_s \sin \theta \cos \phi \\
Z^2 &=r_s \sin \theta \sin \phi \\
Z^3 &=r_s \cos \theta \\
Z^4 &=\frac{1}{\omega} \cosh (\omega v)- r_s \sinh (\omega r_s) 
\end{aligned}$ &
 $\begin{aligned}
Z^0 &=\frac{1}{\omega} \sinh (\omega u)+ r_s \cosh (\omega r_s) \\
Z^1 &=r_s \sin \theta \cos \phi \\
Z^2 &=r_s \sin \theta \sin \phi \\
Z^3 &=r_s \cos \theta \\
Z^4 &=\frac{1}{\omega} \cosh (\omega u)+ r_s \sinh (\omega r_s) 
\end{aligned}$ \\ 

Refs. & \cite[rel.(19), missing $dr_s^2$]{spradlin2001houches} &  \\

metric & $ds^2= -(1-\omega^2 r_s^2)dv^2+ 2dv dr_s - $ & $ds^2= -(1-\omega^2 r_s^2)du^2- 2du dr_s - $ \\ 
 & $ -2 dr_s^2 - r_s^2 d\Omega_2^2$ & $-2 dr_s^2 - r_s^2 d\Omega_2^2$ \\ 

\end{longtable}

\begin{longtable}[h]{ccc}
   \caption{{\bf Kruskal-like charts}}\\
     & $\{u,v,\theta,\phi\}$ &  $\{U,V,\theta,\phi\}$ \\ \hline
introduced: & & Gibbons, Hawking (1977) \cite{gibbons1977cosmological}\\   

			 def. & $\begin{aligned}
Z^0 &=\frac{1}{\omega} \frac{\sinh \frac{\omega (v+u)}{2}}{\cosh \frac{\omega (v-u)}{2}} \\
Z^1 &=\frac{1}{\omega} \tanh \frac{\omega (v-u)}{2} \sin \theta \cos \phi \\
Z^2 &=\frac{1}{\omega} \tanh \frac{\omega (v-u)}{2} \sin \theta \sin \phi \\
Z^3 &=\frac{1}{\omega} \tanh \frac{\omega (v-u)}{2} \cos \theta \\
Z^4 &=\frac{1}{\omega} \frac{\cosh \frac{\omega (v+u)}{2}}{\cosh \frac{\omega (v-u)}{2}} 
\end{aligned}$ &
 $\begin{aligned}
Z^0 &=\frac{U+V}{\omega^2 UV-1} \\
Z^1 &=-\frac{1}{\omega} \frac{\omega^2 UV+1}{\omega^2 UV-1} \sin \theta \cos \phi \\
Z^2 &=-\frac{1}{\omega} \frac{\omega^2 UV+1}{\omega^2 UV-1}  \sin \theta \sin \phi \\
Z^3 &=-\frac{1}{\omega} \frac{\omega^2 UV+1}{\omega^2 UV-1}  \cos \theta \\
Z^4 &=\frac{U-V}{\omega^2 UV-1}
\end{aligned}$ \\ 

metric & $ds^2= -\frac{dudv}{\cosh^2 \frac{\omega (v-u)}{2}} + \tanh^2 \frac{\omega (v-u)}{2}  d\Omega_2^2$ & $ds^2= \frac{2dUdV}{(\omega^2 UV-1)^2} -\frac{1}{\omega^2}\frac{(\omega^2 UV+1)^2}{(\omega^2 UV-1)^2} d\Omega_2^2$ \\ 
Refs. & \cite[rel.(21)]{spradlin2001houches} \cite[rel. (3.8)]{eriksen1995sitter}& \cite[rel.(24)]{spradlin2001houches} \\

& & \\ & & \\
\end{longtable}

\begin{longtable}[ht]{cc}
  \caption{{\bf Constant curvature coordinates}}\\ \label{ccc}
     & $\{\hat{U},\hat{V},{\hat{\xi}},\bar{\hat{\xi}}\}$ \\ \hline
		
 def. & 
 $\begin{aligned}
Z^0 &=\frac{1}{\sqrt{2}} \frac{\hat{U}+\hat{V}}{1-\frac{\omega^2}{2}(\hat{U}\hat{V}-\hat{\xi} \bar{\hat{\xi}})}\\
Z^1 &=\frac{1}{\sqrt{2}} \frac{\hat{V}-\hat{U}}{1-\frac{\omega^2}{2}(\hat{U}\hat{V}-\hat{\xi} \hat{\bar{\xi}})}\\
Z^2 &=\frac{1}{\sqrt{2}} \frac{\hat{\xi}+\bar{\hat{xi}}}{1-\frac{\omega^2}{2}(\hat{U}\hat{V}-\hat{\xi} \bar{\hat{\xi}})}\\
Z^3 &=-\frac{i}{\sqrt{2}} \frac{\hat{\xi}-\bar{\hat{xi}}}{1-\frac{\omega^2}{2}(\hat{U}\hat{V}-\hat{\xi} \bar{\hat{\xi}})}\\
Z^4 &=\frac{1}{\omega} \frac{1+\frac{\omega^2}{2}(\hat{U}\hat{V}-\hat{\xi} \bar{\hat{\xi}})}{1-\frac{\omega^2}{2}(\hat{U}\hat{V}-\hat{\xi} \bar{\hat{\xi}})}
\end{aligned}$ \\ 

ranges & $\hat{U},\hat{V} \in \mathbb{R}$, ${\hat{\xi}} \in \mathbb{C}$ \\
cover &  entire manifold, except singular surface where $\hat{U}=\infty$, $\hat{V}=\infty$ \\

Refs. & \cite[p. 51-52]{griffiths2009exact}  \\

line elem. & $ds^2=\frac{-2d\hat{U}\hat{V}+2\hat{\xi}\bar{\hat{\xi}}}{\left( 1-\frac{\omega^2}{2}(\hat{U}\hat{V}-\hat{\xi} \bar{\hat{\xi}}) \right)^2}$ \\ 
$\sqrt{|g|}$& $ \frac{1}{\left( 1-\frac{\omega^2}{2}(\hat{U}\hat{V}-\hat{\xi} \bar{\hat{\xi}}) \right)^4}$  \\ 

\end{longtable}

\restoregeometry

\end{document}